\documentclass[11pt]{JHEP}

\usepackage{epsfig}
\usepackage{amsmath}
\usepackage{amssymb,amsfonts}
\usepackage{graphicx}

\newcommand\fverb{\setbox\pippobox=\hbox\bgroup\verb}
\newcommand\fverbdo{\egroup\medskip\noindent%
            \fbox{\unhbox\pippobox}\ }
\newcommand\fverbit{\egroup\item[\fbox{\unhbox\pippobox}]}
\newbox\pippobox

\newcommand{\be}{\begin{eqnarray}}
\newcommand{\ee}{\end{eqnarray}}
\renewcommand{\lambda}{\l}

\renewcommand{\t}{\tau}

\newcommand{\R}{\mathbb{R}}
\newcommand{\Co}{\mathbb{C}}

\newcommand{\CP}{\mathbb{C}\mathbb{P}}
\newcommand{\RP}{\mathbb{R}\mathbb{P}}
\newcommand{\Z}{\mathbb{Z}}

\title{Penrose Limits of Orbifolds and Orientifolds }

\author{Emmanuel Floratos$^{1,2}$ and Alex Kehagias$^{1,3}$   \\
  $^1$Institute of Nuclear Physics, N.C.R.P.S. Democritos,
GR-15310, Athens, GREECE\\
$^2$Nuclear and Particle Physics Sector, Univ. of Athens,
GR-15771 Athens, GREECE\\
$^3$Physics Department, NTUA, GR-15773 Zografou, Athens, GREECE\\
\vskip .1in
E-mail: \email{manolis@mail.democritos.gr}, \email{kehagias@mail.cern.ch}}

\preprint{\hepth{0203134}}

\abstract{We study the Penrose limit of various $AdS_p\times S^q$ orbifolds.
The limiting spaces are waves with parallel rays and singular wave fronts.
In particular, we  consider the orbifolds  $AdS_3\times S^3/\Gamma$,
$AdS_5\times S^5/\Gamma$ and $AdS_{4,7}\times
S^{7,4}/\Gamma$ where $\Gamma$ acts on the sphere and/or the $AdS$ factor.
In the pp-wave limit  the wave fronts are the orbifolds  $\Co^2/\Gamma$,
$\Co^4/\Gamma$ and $\R\times \Co^4/\Gamma$, respectively.
When desingularization is possible, we get asymptotically locally pp-wave
backgrounds (ALpp).
The Penrose limit of  orientifolds are also discussed. 
In the $AdS_5\times \RP^5$ case,
 the limiting singularity can be resolved by an
Eguchi-Hanson gravitational instanton. The pp-wave limit of
D3-branes near singularities in F-theory is also presented. Finally, we give
 the embedding of D-dimensional pp-waves in flat $M^{2,D}$ space.
}

\begin{document}

\maketitle 

\setlength{\baselineskip}{1.2\baselineskip}

\section{Introduction}

Plane wave backgrounds with parallel rays (pp-waves) are known to
be exact string solutions \cite{Amati:1988ww},\cite{Horowitz:bv}  . 
 Such backgrounds
have reconsidered recently 
\cite{Blau:2001ne},\cite{Blau:2002dy},\cite{Blau:2002rg}
since they share the similar properties
with flat and $AdS_p\times S^q$ spaces (e.g. maximal supersymmetry). 
The pp-wave background we will
discuss can be considered as the Penrose limit \cite{Penrose}
of $AdS$ spaces 
\cite{Blau:2001ne},\cite{Blau:2002dy},\cite{Blau:2002rg},\cite{Gueven:2000ru},\cite{Berenstein:2002jq}.
The limit is particular and amounts in boosting 
around maximal cycles in $S^q$ and then flatten the resulting
background
 by taking the
standard large radius limit of both $AdS_p$ and $S^q$. 
This limit 
has also a well defined action on the superalgebras of $AdS_p\times S^q$. The
latter are members of
infinite sequence of superalgebras usually denoted by $OSp(N|M)$
and $SU(N|M)$ (There is also the isolated $F(4)$ superalgebra
corresponding to $AdS_6$) \cite{Nahm:1977tg}. 
The superalgebras of the limiting pp-wave backgrounds of 
$AdS_p\times S^q$ can be constructed by contraction of 
the of $OSp(N|M)$ or $SU(N|M)$. An
explicit demonstration  of this has been carried out in \cite{Hatsuda:2002xp}
for  the $SU(2,2|4)$ superalgebra.  

The same limits can also be considered in
 11D $AdS_{4,7}\times S^{7,4}$ M-theory backgrounds. Here, the
resulting pp-wave has been constructed in the past 
\cite{Kowalski-Glikman:wv}
and has recently be recovered \cite{Blau:2001ne}.
An interesting  aspects of all these constructions is that the
Penrose limit  preserves all
supersymmetries of the original background. Thus, the resulting
 pp-waves are
maximally supersymmetric. The fact that maximally supersymmetric
space-times have
played a central role in understanding string and M-theory
explains in part the recent interest in these pp-wave backgrounds. 
Additional support for studying them, came from  the recent
proposal of Berestein, Maldacena and
Nastase (BMN)  \cite{Berenstein:2002jq} 
 that the string spectrum on a pp-wave background  arises
from the large $N$ limit of ${\cal{N}}=4$ SYM theory in 4d. This
has been demonstrated by summing a subset of planar diagrams and
it is a very interesting extension of the original AdS/CFT
correspondence as it involves massive string modes. The large
$N$ limit that has been employed is not just large 't Hooft
coupling $g_{YM}^2N\to \infty$ but also fixed $g_{eff}^2=g_{YM}^2N/J^2$
($J$ is a global charge).
Then, correlation functions for operators of scaling dimension $\Delta$
are calculated in large 't Hooft coupling, fixed $g_{eff}$ and
finite $\Delta\!-\!J$.

Here, we  consider various orbifolds of $AdS_p\times S^q$
and their pp-wave limit. As these orbifolds have less
supersymmetry, the corresponding pp-waves will also have less
supersymmerties provided that the singularities are not washed out in
the limit. The spectrum on these backgrounds
should then arise from the corresponding SCFTs in the BMN limit. 
The same can be done for
$AdS$ orientifolds and we present the pp-wave limits of the near
horizon
of D3-branes on  O3-plane as well as in F-theory (at constant
coupling)
\cite{Sen:1996vd},\cite{Banks:1996nj}, \cite{Dasgupta:1996ij}.

This paper is organized as follows. In section 2, we describe the
Penrose limits of   $AdS_p\times S^q$ backgrounds which occur in string and
M-theory. In section 3 we construct various orbifolds of the
backgrounds presented in the previous section and we discuss their
possible desingularizations. In section 4, we present the pp-wave
limit of $AdS_5\times S^5$ orientifold. Finally, in appendix A we
give the embedding of the D-dimensional pp-wave in flat
$M^{2,D-1}$ space, while in appendix B we describe the ALE space
$\Co^3/\Z_3$ we used as an example in section 3.

\section{Penrose limit of $AdS_p\times S^q$}

$AdS$ spaces arise in many instances in string and M-theory. In fact, 
supersymmetric $AdS_p$ backgrounds are possible, according to 
\cite{Nahm:1977tg}, for $p=2,3,4,5,6,7$. They appear as vacuum of 10
and 11 dimensional supergravity as well as the near horizon limit of
D-brane and M-brane configurations  (See \cite{Aharony:1999ti} for a
review). 
Keeping the discussion as general
as possible, 
let us recall the Penrose limit of $AdS_p\times S^q$, consider first
in \cite{Blau:2002dy}. The  metric
of $AdS_p\times S^q$ is
\be
ds^2=R_{A}^2\left(d\rho^2+\sinh^2\!\rho \, d\Omega_{p-2}^2-
\cosh^2\!\rho\,
dt^2\right)+R_S^2\left(d\theta^2+\sin^2\!\theta\, 
d\Omega_{q-2}^2+\cos^2\!\theta
\,d\psi^2\right)\, , \label{p1}
\ee
where $R_A,R_S$ are the radius of $AdS_p$ and $S^q$, respectively.
We will consider  the limiting geometry seen by a
fast moving observer in the $\psi$ direction at $\rho=0,\theta=0$.
For this, we introduce the coordinates
\be
x^+={t+\alpha \psi\over 2}\, , ~~~ x^-=R_A^2\left({t-\alpha
\psi\over 2}\right)\, , ~~~ x=R_A \rho\, , ~~~ y=R_S\theta\, ,
~~~\alpha={R_S\over R_A}\label{pen}
\ee
in the metric (\ref{p1})
and then take the limit $R_{A,S}\to \infty$,  keeping
$\alpha$ finite. The resulting space-time is non singular and its 
metric is
\be
ds_P^2=-4 dx^+dx^--(x^2+\alpha^{-2}
y^2)dx^+{}^2+dx^2+x^2d\Omega_{p-2}^2+dy^2+y^2d\Omega_{q-2}^2\, .
\ee

Next, we will examine each possible $AdS$-geometries appear in
string and M-theory (except the $AdS_2$ and $AdS_6$ cases).
\vskip .1in

\subsection{$AdS_3\times S^3$}

This background appears in the near horizon geometry of
the D1/D5 system \cite{Horowitz:1996ay}. The 10D geometry is 
$AdS_3\times S^3\times
M^4$ and the D5-branes are wrapped on $M^4$. In order that
supersymmetry 
is preserved,
$M^4$ is either $T^4$ or $K^3$ for 8 or 4 surviving supersymetries.
Omitting the irrelevant for the discussion 
 $M^4$ factor, the bosonic symmetry of this background is
$SO(2,2)\times SO(4)$ which is the bosonic part of the 2D SCFT living in the
boundary of $AdS_3$ \cite{Maldacena:1997re}. The metric is
\be
ds^2=R_{A}^2\left(d\rho^2+\sinh^2\!\rho \, d\phi^2-
\cosh^2\!\rho\,
dt^2\right)+R_S^2\left(d\theta^2+\sin^2\!\theta\, d\chi^2+\cos^2\!\theta
\,d\psi^2\right)\, , \label{pp1}
\ee
with $R_A=R_S$. Defining the coordinates as in  (\ref{pen}) 
and taking the $R_A\to \infty$
limit, we end up with the
pp-wave
\be
&&ds_P^2=-4 dx^+dx^--\mu^2(x^2+
y^2)dx^+{}^2+dx^2+x^2d\phi^2+dy^2+y^2d\chi^2\nonumber \\&&
H_{+34}=H_{+56}=2\mu \, , \label{ads3}
\ee
where $H$ is the RR three-form of the D1/D5 system. It is evident
that the pp-wave front $x^+=$const. is flat 4D Euclidean space.

\subsection{$AdS_5\times
S^5$}

 The  $AdS_5\times S^5$ background is by far the most celebrated
one. It is realized as the near horizon limit of $N$ coincident
D3-branes and it is maximally supersymmetric (32 supersymmetries). 
$AdS_5\times S^5$ with 
$N$ units of five-form
flux is conjectured to be  the supergravity dual
of $SU(N)$ ${\cal N}=4$ gauge theory at large $N$ 
\cite{Maldacena:1997re}. The bosonic
symmetry of the background is $SO(4,2)\times SO(6)$. The $SO(4,2)$
part is realized as the conformal symmetry of the $SU(N)$
${\cal{N}}$ gauge theory while  $SO(6)$ is the R-symmetry.
Starting form the $AdS_5\times S^5$ metric
\be
ds^2=R^2\left(d\rho^2+\sinh^2\!\rho \, d\Omega_3^2-
\cosh^2\!\rho\,
dt^2+d\theta^2+\sin^2\!\theta\, d\Omega_3^2+\cos^2\!\theta
\,d\psi^2\right) \, , 
\ee
the geometry seen by a fast moving
observer in the $\psi$ direction sitting at $\rho=0, ~\theta=0$
can be obtained, according to eq.(\ref{pen}), by introducing the
coordinates
\be
x^+={t+\psi\over 2}\, , ~~~x^-=R^2\left({t-\psi\over 2}\right)\, , ~~~
x=R \rho\, , ~~~ y=R\theta\, , \label{pen5}
\ee
and then take the limit $R\to \infty$. The resulting non-singular
space-time  metric and the surviving RR self-dual five-form
are \footnote{This is exactly the metric on the group manifold of the 
10D Heisenberg group \cite{Kehagias:1994iy}. It is a generalization of
the Nappi-Witten background \cite{Nappi:1993ie}
with RR fields instead of NS/NS
antisymmetric 2-form. It can also be considered as a particular
contruction of a WZW model \cite{Sfetsos:1994vz}. }
\be
&&ds^2=-4dx^+dx^--\mu^2{\vec{r}}{~}^2{dx^+}^2+d{\vec{r}}{~}^2\,
, \nonumber \\ &&F_{+1234}=F_{+5678}=4\mu \label{pp4}
 \ee where
${\vec{r}}{~}^2={\vec{x}}{~}^2+{\vec{y}}{~}^2$. Clearly, the wave
front is flat 8D Euclidean space. The type IIB superstring in the
above background has been
argued to be exactly solvable, described by free massive fields  
\cite{Metsaev:2001bj}, \cite{Metsaev:2002re}, \cite{Russo:2002rq}. 

\vskip .2in
\subsection{$AdS_4\times S^7$}
\vskip .1in
 This geometry appears in the near horizon of $N$ coincident M2-branes
in M-theory. It is the supergravity dual of ${\cal{N}}=8$ 3D SCFT living on
the 
M2 worldvolume.
The bosonic symmetry of the supergravity background is $SO(3,2)\times
SO(8)$. 
As usual, the 
$SO(3,2)$ is identified with the conformal symmetry of the boundary
 SCFT while the $SO(8)$ 
part is the R-symmetry group.
 Starting form the $AdS_4\times S^7$ metric
\be
ds^2=R_A^2\left(d\rho^2+\sinh^2\!\rho \, d\Omega_2^2-
\cosh^2\!\rho\, dt^2\right)+R_S^2\left(d\theta^2+\sin^2\!\theta\,
d\Omega_5^2+\cos^2\!\theta \, d\psi^2\right)\, ,  \label{pp47}
\ee where $R_S=2R_A$,
the Penrose limit  is obtained by defining
\be
x^+={1\over 2}\left(t+ 2 \psi\right)\, , ~~~
x^-={R_A^2\over 2}\left(t-2 \psi\right)\, , ~~~
x=R_A \rho\, , ~~~ y=2 R_A\theta\, , \label{pen4}
\ee
and letting $R_A\to \infty$. The resulting space-time
metric and 4-form are
 \be
&&ds_P^2=-4 dx^+dx^--\mu^2(x^2+{1\over 4}
y^2)dx^+{}^2+dx^2+x^2d\Omega_{2}^2+dy^2+y^2d\Omega_{5}^2 \, ,
\nonumber \\
&&F_4=3\mu \, dx^+\wedge dx^1\wedge dx^2\wedge dx^3 \label{ppp4}
\ee
so that the near horizon of M2-branes has been 
turned into  pp-waves with flat wave fronts.

\vskip .2in
\subsection{$AdS_7\times S^4$}
\vskip .1in
Similarly to the above, this geometry is realized in the near
horizon limit of $N$ coincident M5 branes in M-theory. They
describe the supergravity duals of large $N$ 6D SCFT.  
The bosonic symmetry of this background is
$SO(6,2)\times SO(5)$ which is the bosonic part of the 6D SCFT. The metric
of $AdS_7\times S^4$ is
\be
ds^2=R_A^2\left(d\rho^2+\sinh^2\!\rho \, d\Omega_5^2-
\cosh^2\!\rho\,
dt^2\right)+R_S^2\left(d\theta^2+\sin^2\!\theta\, d\Omega_2^2+\cos^2\!\theta
\, d\psi^2\right)\, . \label{ads77}
\ee
Here, $R_A=2R_S$ and the limit $R_A\to \infty$ after the
transformation
\be
x^+={1\over 2}\left(t+ {\psi\over 2}\right)\, , ~~~
x^-={R_A^2\over 2}\left(t-{\psi\over 2}\right)\, , ~~~
x=R_A \rho\, , ~~~ y= {R_A\over 2}\theta\, , \label{pen7}
\ee
leads to the pp-wave background and 4-form
\be
&&ds_P^2=-4 dx^+dx^--(x^2+ 4
y^2)dx^+{}^2+dx^2+x^2d\Omega_{5}^2+dy^2+y^2d\Omega_{2}^2 \, ,
\nonumber \\
&&F_4=6\mu\,  dx^+\wedge dy^1\wedge dy^2\wedge dy^3\, . \label{pp7}
\ee
In fact, the pp-wave limits of both $AdS_4\times S^7$ and
$AdS^7\times S^4$ in Eqs.(\ref{ppp4}) and (\ref{pp7}),
respectively, are the same (as can be seen by the transformation
$x\leftrightarrow y, ~ x^+\to x^+/2$) and the two maximally supersymmetric
backgrounds (\ref{pp47}) and  (\ref{ads77}) have the unique maximally 
supersymmetric pp-wave background (\ref{ppp4}) (or (\ref{pp7})) 
found in  \cite{Kowalski-Glikman:wv}.

\section{Penrose limits of $AdS_p\times S^q$ orbifolds}

Branes can be put at conifold \cite{Kehagias:1998gn}, 
\cite{Klebanov:1998hh}, \cite{Morrison:1998cs}
or orbifold singularities \cite{Douglas:1996sw},
\cite{Douglas:1997de}, \cite{Kachru:1998ys}. The near
horizon limit of such configurations lead to either singular or
non-singular spaces. Here we will examine orbifolds of $AdS_p\times
S^q$ and we will find their pp-wave limit. The corresponding limit 
at the conifold has been studied in \cite{Itzhaki:2002kh}, 
\cite{Gomis:2002km}, \cite{Zayas:2002rx}. 
 The type of orbifold
theories we will consider are such that they lead to singularities
on the wave front of the pp-wave. Some of these singularities may
be resolved to a smooth Ricci flat space (leading  not to flat wave fronts
 but rather to  Ricci flat ones) and some not. We will discuss each case of
$AdS$ background
separately.

\subsection{$AdS_3\times S^3$ orbifolds}

This background may be embedded in 8D space-time $M^{2,6}$ with
metric (in an obvious complex notation)
\be
ds_8^2=-|dZ_0|^2+|dZ_1|+|dZ_2|^2+|dZ_3|^2\, , 
\ee
as the hypersurface
\be
|Z_0|^2-|Z_1|^2=R^2\, , ~~~~|Z_2|^2+|Z_3|^2=R^2\, . \label{zz}
\ee
The parametrization
\be
Z_0=R\cosh\rho e^{it}\, ~~~Z_1=R\sinh\rho e^{i\phi}\, , ~~~
Z_2=R\cos\theta e^{i\psi}\, ~~~Z_3=R\sin\theta e^{i\chi}\, ,
\ee
leads directly to the metric (\ref{pp1}).
Many orbifolds of the hypersurface (\ref{zz}) may be taken. For
example, we may consider the $\Z_k$ action
$$
Z_3\to e^{2\pi i /k}Z_3\, , 
$$
which has fixed points the cycle $|Z_2|^2=R^2$. The pp-wave limit
is given again by eq.(\ref{ads3}). The only difference is the
periodicity of $\chi$ which instead being $2\pi$ is now $2\pi/k$.
As a result, the wave front has been turned from $\Co^2$ in the
$AdS_3\times S^3$, to $\Co\times \Co/\Z_k$ in the $AdS_3\times S^3/\Z_k$
case. There exist a conical singularity (at $y=0$ in (\ref{ads3}))
which, however, cannot be resolved. An example, where the
singularity can be resolved is provided by the orbifold action
\be
\Z_k~~: Z_3\to e^{2\pi i /k}Z_3\, ,
~~~Z_1\to e^{-2\pi i /k}Z_1\, . \label{zz3}
\ee
Here,  the pp-wave limit is  again given by the metric in
(\ref{ads3}), where now  the periodicities of both  $\phi,\chi$
are $2\pi/k$. (Orbifolds of  the $AdS$ factor has been considered in 
AdS/CFT correspondence in \cite{Horowitz:1998xk}, \cite{Gao:1999er}).
In fact, the action
(\ref{zz3})
 identifies $(\phi,\chi)\equiv
(\phi+2\pi/k,\chi+2\pi/k)$.
Thus, the wave front is $\Co^2/Z_k$, which can
be resolved by an ALE space.

Replacing $\Co^2/\Gamma$ by an ALE
space is not the end of the story. We should also make sure that
supergravity equations are satisfied.
We will do this below  for
the  general case by
 looking for background with metric and
three-form RR field of the form
\be
&&
ds^2=-4 dx^+dx^--S(x^i){dx^+}^2+g_{ij}dx^i dx^i\, , 
~~~i,j=1,...,4\, , \nonumber \\
&&H_{+ij}={1\over 2}\epsilon_{ijmn}{H_+}^{mn}\, , 
\ee
where $g_{ij}$ is the metric of a 4D-dimensional wave-front space
$K^4$. The Ricci tensor is
\be
R_{++}={1\over 2} \nabla^i\nabla_iS\, , ~~~R_{ij}=R_{ij}(g)\, , 
\ee
so that the Einstein equations
\be
R_{MN}={1\over 4} \left(H_{MKL}{H_N}^{KL}-{1\over 12 }
H^2g_{MN}\right)\, , 
\ee
reduce to
\be
&&{1\over 2} \nabla^i\nabla_iS={1\over 4} H_{+mn}{H_+}^{mn}\, , 
\nonumber \\
&& R_{ij}(g)=0\, . 
\ee
In the Penrose limit of $(AdS_3\times S^3)/\Z_k$ with $\Z_k$ as 
in eq.(\ref{zz3}), $H_{MNP}$ is
given by eq.(\ref{ads3}) so that
\be
ds^2=-4 dx^+dx^--\mu^2S(x){dx^+}^2 +
g_{ij}dx^idx^j\, , 
\ee
where $g_{ij}$ is the metric of the  ALE space.  $S$
is determined by the equation
\be
\nabla^i\nabla_i S=8 \, .\label{se}
\ee
In the case of an 
$AdS_3\times S^3/\Z_2$ orbifold, the singular $\Co/\Z_2$ can be
replaced by the Eguchi-Hanson gravitational instanton with metric
\be
ds_{EH}^2={dr^2\over 1-{a^4\over
r^4}}+r^2(\sigma_1^2+\sigma_2^2)+r^2\left(1-{a^4\over
r^4}\right)\sigma_3^2\, .
\ee
The solution to eq.(\ref{se}) for the EH metric then leads to the
pp-wave limit of $(AdS_3\times S^3)/\Z_2$
\be
ds^2=-4 dx^+dx^--\mu^2\left(r^2+{a^2\over 2}
\ln\left(r^2-a^2\over r^2+a^2\right)\right)){dx^+}^2 +
ds_{EH}^2 \, .\label{ehh}
\ee
Clearly, at $r\to \infty$, the wave fronts of  (\ref{ehh}) 
become $\Co^2/\Z_k$ as it should. 
In the case  of $\Z_k\subset SU(2)\, , (k>2)$,
$\Co^2/\Z_k$ should be replaced by the
Gibbons-Hawking multi-center gravitational instanton with metric
\be
ds^2_{GH}=V(d\tau+\vec{\omega}\cdot d\vec{x})^2+V^{-1}d\vec{x}{}^2\, ,
\ee
where
\be
V^{-1}=\sum_{i=1}^k{1\over |\vec{x}-\vec{x}_i|}\, , \ ~~~\vec{\nabla} \times
\vec{\omega}=\vec{\nabla} V^{-1}\, .
\ee
Then, the solution of eq.(\ref{se}) will formally be given as
\be
S=-{1\over \pi} \sum_{i=1}^k\int { d^3x' \over
|\vec{x}-\vec{x}'||\vec{x}'-\vec{x}_i|}\, .
\ee

It is clear, that other orbifolds $(AdS_3\times S^3)/\Gamma$ 
may be considered,
where $\Gamma\subset SU(2)$ in order to preserve supersymmetry
(such orbifolds preserve half the supersymmetry  $AdS_3\times
S^3$ preserves).
Their pp-wave limits have then the singular wave fronts
$\Co^2/\Gamma$ which can be made smooth by replacing them (after
appropriate modification of the metric as shown above) by an ALE
space. Such spaces may be called ALpp spaces as they are
asymptotically (in the wave-front sense) locally pp-waves.

\subsection{Orbifolds of $AdS_5\times S^5$}

Similarly to   $AdS_3\times S^3$, $AdS_5\times S^5$ can be embedded in
12D  flat space-time $M^{2,10}$
with metric
\be
ds_{12}^2=-dX_0^2+dX_1^2+...+dX_{10}^2-dX_{11}^2\, , \label{mm}
\ee
as the hypersurface
\be
&&X_0^2-X_1^2-X_2^2-X_3^2-X_4^2+X_{11}^2=R^2\, , \label{ads} \\&&
|Z_1|^2+|Z_2|^2+|Z_3|^2=R^2\, , \label{par} 
\ee 
where $Z_1=X_5+iX_6,~Z_2=X_7+iX_8, ~Z_3=X_9+iX_{10}$.
The parametrization of (\ref{ads})
\be
&&X_0=R\cosh\rho\, \cos t \, \, ~~~~X_{11}=R\cosh\rho\, \sin t\nonumber \\
&&
X_a=R\sinh\rho \Omega_a\, , ~~~~(a=1,...,4,
~~~\sum_a\Omega_a^2=1)\, , 
\ee
with $0\leq \rho<\infty$ and $0\leq t< 2\pi$ covers the whole of
the (\ref{ads}) hyperboloid. Together with the angular coordinates
$\Omega_a$, which parametrize a unit $S^3$, are the global
coordinates of the $AdS_5$. Similarly, the $S^5$ can be
parametrized as
\be
&& Z_1=R \cos\theta\,e^{i\psi}\,~,~~~ Z_2=R \sin\theta\, U_1\,,
~~~\nonumber \\&&
Z_3=R \sin\theta\, U_2\,, ~~~~|U_1|^2+|U_2|^2=1 \, , \label{s5}
\ee
where $0\leq \theta < \pi/2$, $0\leq \psi<2\pi$ and the complex
$U_{1,2}$ form a unit  $S^3$.
We may act now
with a discrete subgroup $\Gamma$ of the isometry group 
$SO(6)\sim SU(4)$ of $S^5$.
There are two distinct
cases which preserve supersymmetry \cite{Kachru:1998ys},
\cite{Lawrence:1998ja}. The first one is when 
$\Gamma\subset SU(3)\subset SU(4)$ and
leads to ${\cal{N}}=1$ supersymmetric $SU(N)$ gauge theory.
The second case is when $\Gamma\subset SU(2)\subset SU(4)$ with
${\cal{N}}=2$ supersymmetry. Both cases have the same Penrose limit.
For concreteness, we will consider a  $\Gamma \subset SU(2)$
acting
on the coordinates $Z_2,Z_3$. The $\Gamma$ action  fixes
the $S^1$ in $S^5$
\be
|Z_1|^2=1\, \ ~~ Z_2=Z_3=0\, , \label{s1}
\ee
 and thus $S^5/\Gamma$ is a  singular orbifold. In the parametrization
 (\ref{s5}),
 $\Gamma$ acts freely on $S^3$ forming
the  space $S^3/\Gamma$. The latter degenerates at $\theta=0$
fixing the singular $S^1$ in eq.(\ref{s1}) with coordinate $\psi$. 
The Penrose
limit (\ref{pen}) can now be taken leading to a pp-wave background
\be
&&ds^2=-4dx^+dx^--\mu^2({\vec{x}}{~}^2+{\vec{y}}{~}^2){dx^+}^2+
d{\vec{x}}{~}^2\,+
dy^2+y^2d\tilde{\Omega}_{ 3}^2
,\nonumber  \\ &&F_{+1234}=F_{+5678}=4\mu \, , \label{pps}
 \ee
 where $d\tilde{\Omega}_{3}^2$ is the metric on
$S^3/\Gamma$.
 There exists a singularity at $y=0$ and 
the Penrose limit of the $AdS_5\times S_5/\Gamma$ produces the
singular wave fronts 
 $\Co^2\times \Co^2/\Gamma$. As  orbifolding by $\Gamma$ breaks half
the supersymmetries, the Penrose limit of $AdS_5\times S_5/\Gamma$
also breaks half, leading to 16 surviving supersymmetries.
We may blow up the singularity at $y=0$ by replacing $\Co^2/\Gamma$ by
an ALE space in such a way that the supergravity equations are
still satisfied.
Thus, we should look for background with metric and
five-form RR field of the form
\be
&&
ds^2=-2 dx^+dx^--S(x^i){dx^+}^2+g_{ij}dx^i dx^i\, , ~~~i,j=1,...,8\, ,
\nonumber \\
&&F_{+ijk\ell}={1\over 4!}\epsilon_{ijk\ell mnpq}{F_+}^{mnpq}\, , 
\ee
where $g_{ij}$ is the metric of an 8D-dimensional wave-front $K^8$.
The Einstein equations
\be
R_{MN}={1\over 96} F_{MKLPQ}{F_N}^{KLPQ}\, , 
\ee
reduce to
\be
&&{1\over 2} \nabla^i\nabla_iS={1\over 96} F_{+mnpq}{F_+}^{mnpq}\, , 
\nonumber \\
&& R_{ij}(g)=0\, .
\ee
In the Penrose limit of $AdS_5\times S^5/\Gamma$, $F_{MKLPQ}$ is
still
given by eq.(\ref{pp4}) so that
\be
ds^2=-4 dx^+dx^--\mu^2({\vec{x}}{~}^2+S(y)){dx^+}^2+d{\vec{x}}{~}^2 +
g_{\mu\nu}dy^\mu dy^\nu
\ee
where $g_{\mu\nu}$ is the metric on the blow-up $\Co^2\times \Co^2/\Gamma$ and $S$
satisfies again eq.(\ref{se}).


We may also consider orbifold of the $AdS_5$ space
\cite{Horowitz:1998xk},
\cite{Gao:1999er}. For
example, with $\zeta_0=X_0+i X_{11},~\zeta_1=X_1+iX_2,
~\zeta_2=x_3+i X_4$, the $AdS$ hyperboloid (\ref{ads}) is written
as
\be
|\zeta_0|^2-|\zeta_1|^2-|\zeta_2|^2=R^2
\ee
 The $\zeta$'s are expressed as
\be
&&\zeta_0=R \cosh\rho\, e^{i t}\, , ~~~\zeta_1=R \sinh\rho\, V_1\, ,
~~~\zeta_2=R \sinh\rho \, V_2\, ,\nonumber \\&&
|V_1|^2+|V_2|^2=1 \, \label{vv}
\ee
where $V_{1,2}$ parametrize an $S^3$. We choose a subgroup $\Gamma'\subset SU(2)$
which acts on $\zeta_1,\zeta_2$. Then $\Gamma'$ acts freely on the
$S^3$ of (\ref{vv}).  The fixed point set of $\Gamma'$ on $AdS_5$ is
 the singular $S^1$
$$|\zeta_0|^2=R^2\, , ~~~~ \zeta_1=\zeta_2=0$$
The Penrose limit can similarly be taken leading to the singular
pp-wave background with wave front $\Co^2/\Gamma'\times \Co^2/\Gamma$
\be
&&ds^2\!=\!-4dx^+dx^-\!-\!\mu^2({\vec{x}}\!{~}^2\!+\!{\vec{y}}\!{~}^2)
{dx^+}^2+dx^2+x^2d\tilde{\Omega}_3^{'2}+
dy^2+y^2d\tilde{\Omega}_3^2\, , 
 \label{ppps} \\ &&F_{+1234}=F_{+5678}=\mu\, , 
 \ee
where $\tilde{\Omega}_3^{2},~~\tilde{\Omega}_3^{'2}$ are the metrics
on $S^3/{\Gamma}$ and $S^3/\Gamma'$, respectively.
Again, the singularities may be blown up by replacing $\Co^2/\Gamma'\times
\Co^2/\Gamma$ by the two ALE spaces $M^4\times N^4$.
 Then, the desingularized Penrose limit of
$AdS_5/\Z_{k'}\times S^5/\Z_k$ is
\be
ds^2=-4 dx^+dx^--\mu^2\Big{(}S(x)+T(y)\Big{)}{dx^+}^2+ds_M^2(x) +
ds_N^2(y)\, , 
\ee
where $ds_M^2(x)$, $ds_N^2(y)$ are the metrics on $M^4$, $N^4$,
respectively whereas
 $S,T$ satisfy
\be
\nabla_{(M)}^2S=4\, , ~~~\nabla_{(N)}^2T=4\, .
\ee
It is also possible to consider more general orbifolds of
the $AdS_5\times S^5$ geometry. For example consider 
$(AdS_5\times S^5)/\Z_k$ defined by the $\Z_k$ action
\be
\Z_k: ~~~Z_1\to e^{2i\pi/k}Z_1\, , ~~Z_1\to e^{-2i\pi/k}Z_1\, ,
~~\zeta_1\to e^{2i\pi a/k}\zeta_1\, ,
~~\zeta_2\to e^{-2i\pi a/k}\zeta_2\, ,
\ee
where $a,k$ relatively prime. Then, in the pp-wave limit, the
singular 
wave fronts
$(\Co^2\times \Co^2)/\Z_k$ are obtained, 
 where $\Z_k$ as above. It is known that
the singularity in this case is terminal \cite{Morrison}, 
\cite{Morrison:1998cs}. If $a=1\, , 
k=2$ we get
16 supersymmetries (${\cal{N}}=8$) whereas if $a=\pm 1\, , k>2$
there exist only 12 surviving supersymmetries (${\cal{N}}=6$).
Finally, if $a\neq \pm 1$, we end up with 8 supersymmetries
(${\cal{N}}=4$). Neither of these singularities have Calabi-Yau
resolutions.

The Penrose limits of the $AdS_5\times S^5$ orbifolds are by now more or
less clear. Defining the pp-wave limit
 of the $AdS_5\times S^5$ space by the hypersurface in
$M^{2,10}$
\be
&&X_{10}-X_{11}=\left(X_0+X_{9}\right)^2\, ,\\
&&X_{10}+X_{11}={\mu^2\over 8}\left(X_1^2+X_2^2+...+X_8^2\right)\, , 
\ee
as discussed in the appendix, the orbifolds we have considered
above are in fact orbifolds of $\Co^4$ with coordinates
$Z_1,Z_2,\zeta_1,\zeta_2$ defined above. We have only consider some 
special cases of such possible orbifolds. However,
it is natural to assume that there exists a larger class of
$\Co^4$ orbifolds (containing the ones we have discussed above)
which are  wave fronts of  Penrose limits of
associated $AdS_5\times S^5$ orbifolds. By desingularizing the 
$\Co^4$ orbifolds, 
we may obtain smooth pp-wave limits of $AdS_5\times S^5$ orbifolds, where
the locally flat wave fronts are replaced  by  Ricci-flat
ones of appropriate holonomy in order supersymmetry to be preserved.


\subsection{$AdS_4\times S^7$ orbifolds}

As in the previous cases, $AdS_4\times S^7$ can be embedded in
$M^{2,11}$ with metric
\be
ds_{12}^2=-dX_0^2+dX_1^2+...+dX_{10}^2+dX_{11}^2-dX_{12}^2\, , 
\label{mmm} 
\ee 
as the hypersurface
\be
&&X_0^2-X_1^2-X_2^2-X_3^2+X_{12}^2=R^2\, , \label{ads4} \\&&
|Z_1|^2+|Z_2|^2+|Z_3|^2+|Z_4|^2=4 R^2\, . \label{par7} 
\ee 
where $Z_1=X_4+iX_5,~Z_2=X_6+iX_7, ~Z_3=X_8+iX_{9}, ~Z_4=X_{10}+iX_{11}$. The
parametrization of (\ref{ads4})
\be
&&X_0=R\cosh\rho\, \cos t \, , ~~~~X_{12}=R\cosh\rho\, \sin
t\, , \nonumber \\ && X_a=R\sinh\rho \Omega_a\, , ~~~~(a=1,...,3,
~~~\sum_a\Omega_a^2=1)\, ,  \ee
with $0\leq \rho<\infty$ and $0\leq t< 2\pi$ covers the whole of
the (\ref{ads4}) hyperboloid. The angular coordinates
$\Omega_a$  parametrize a unit $S^2$ and together with $\rho,t$ are the global
coordinates of the $AdS_4$. Similarly, the $S^7$ in eq.(\ref{par7}) can be
parametrized as
\be
&& Z_1=R \cos\theta\,e^{i\psi}\,~,~~~ Z_a=R \sin\theta\, U_a\,, 
~~(a=1,2,3)\nonumber
\\&&
|U_1|^2+|U_2|^2+|U_3|^2=1 \, \label{s55}
\ee
where $0\leq \theta < \pi/2$, $0\leq \psi<2\pi$ and the complex
$U_{1,2,3}$ form a unit  $S^5$.

We may now consider orbifolds $AdS_4\times S^7/\Gamma$ where
$\Gamma\subset SU(4)$ as in \cite{Ferrara:1998vf}.  
 Such orbifolds acts freely on the $S^5$ of
eq.(\ref{s55}) but not freely on $AdS_4\times S^7$ as the
hypersurface $\theta=0$ is the fixed point set of  $\Gamma$.
In order that supersymmetry is preserved, we should take
$\Gamma\subset SU(3)\subset SU(4)$ (1/4 supersymmetry) or
$\Gamma\subset SU(2)$ (1/2 supersymmetry). The pp-wave limit of
the above orbifolds is the background (\ref{pp4}) but now the $S^5$
metric $d\Omega_5$ is replaced with the metric on $S^5/\Gamma$. Thus,
the wave fronts are not any more $\R^9$ but $\R^3\times
\Co^3/\Gamma$. We may replace in this case the singular
$\Co^3/\Gamma$ with an ALE space of $SU(3)$ holonomy (for
$\Gamma\subset SU(3)$) getting a smooth pp-wave background. The
metric is then
\be
ds^2=-2 dx^+dx^-\!-\!\mu^2\Big{(}(\vec{x}{~}^2+S(x^i)\Big{)}{dx^+}^2+
d\vec{x}{~}^2+g_{ij}dx^i dx^i\, ,  ~~
i,j=1,...,6\, , 
\ee
where $\vec{x}$ is in $\R^3$, $g_{ij}$ is the metric on 
the ALE space and $S(x^i)$
satisfies
\be
\nabla^i\nabla_i S=3
\ee
In the special case where $\Gamma=Z_3\subset SU(3)$, we may use
the ALE space described in the appendix B. In this case, the
desingularized  pp-wave limit of $AdS_4\times S^7/\Z_3$ turns out
to be
\be
&&ds^2=-2 dx^+dx^--\mu^2\left(x^2+{r^2\over 4}+
{L^2\over 24}\ln\Big{(}
{L^4+r^4-2L^2r^2\over L^4+r^4+L^2r^2}\Big{)}\right.\nonumber \\&&
\phantom{xxxxxxx}+{L^2\over 4 \sqrt{3}}\left.\Big{\{}
\arctan({2r +L\over 2 \sqrt{3}})-
\arctan({2r -L\over 2 \sqrt{3}})\Big{\}}\right)
{dx^+}^2+
d\vec{x}{~}^2+d\sigma^2
\ee
where $d\sigma^2$ is the metric (\ref{SS1}). Clearly, at $r\to \infty$
we recover the $\R^3\times \Co^3/\Z_3$ wave fronts. 

\subsection{$AdS_7\times S^4$ orbifolds}

We may repeat the same procedure as above in the present case as
well. We may embed this background in $M^{2,11}$ as before with the role of
$AdS_4$ and $S^7$ interchanged. There exist only one
orbifold of $S^4$ which preserves supersymmetry. It is the
$S^4/\Z_2$ where $\Z_2$ acts on the $S^2$ with metric
$d\Omega_2^2$ in eq.(\ref{ads77}). However, this orbifold is
singular at $y=0$ and it cannot be resolved. Of course, one may
consider orbifolds of the $AdS_7$ factor but we will not go into
details as this case is similar to the ones already studied above.
We should only mention that although $AdS_4\times S^7$ and
$AdS_7\times S^4$ have the same Penrose limits, their orbifolds do
not share the same property.

\section{pp-waves from $AdS_5\times S^5$ orientifold}

The $AdS_5\times S^5$ orientifold is the near horizon limit of
D3-branes at orientifold planes. The study of D3 branes at
orientifolds is similar to orbifolds (See for example
\cite{Giveon:1998sr}). 
The only difference is the
twisted sector which is absent for orientifolds. We will first
consider the near horizon of D3-branes on an  O3-plane. As the
O3-plane breaks the same supersymmetries with the D3, in the near
horizon we will have again the maximum 32 unbroken
supersymmetries. The near horizon geometry is actually
$AdS_5\times \RP^5$ where $\RP^5=S^5/\Z_2$. The $\Z_2$ acts by
identifying opposite points on the $S^5$ so that there are no
fixed points. As a string goes around a non contractible cycle,
connecting opposite points in $\RP^5$, it reverses its orientation
which is a manifestation of the orientifold projection. There are
two types of the orientifold projection. One leads to $SO(2N)$
${\cal{N}}=4$ theory for $N$ D3-branes on the O3-plane, while the
other leads to an $USp(2N)$ theory. These different sting theories
are implemented in the $AdS_5\times \RP^5$ setup by turning on
$B_{NS-NS}$ 2-form (discrete torsion) in the non-trivial
cohomology class $H^3(\RP^5,\tilde{\Z})=\Z_2$ 
\cite{Witten:1998xy}. The  metric  of
the $AdS_5\times \RP^5$ geometry is
\be
ds^2=R^2\left(d\rho^2+\sinh^2\!\rho \, d\Omega_3^2-
\cosh^2\!\rho\, dt^2+d\theta^2+\sin^2\!\theta\,
d\tilde{\Omega}^2_{3}+\cos^2\!\theta \,d\tilde{\psi}^2\right) \ee
where $0 \leq \tilde{\psi}<\pi$ and $d\tilde{\Omega}^2_{3}$ is the
metic on $\RP^3=S^3/\Z_2$. The Penrose limit is then
\be
&&ds^2=-4dx^+dx^--\mu^2(\vec{x}{}^2+\vec{y}{~}^2){dx^+}^2+d{\vec{x}}{~}^2+
dy^2+y^2 d\tilde{\Omega}^2_{3}\,. \label{ppr}
\ee
 It is clear that
the wave fronts have an $A_1$ singularity and are $\Co^2\times
\Co^2/\Z_2$. The singular $\Co^2/\Z_2$ can be replaced by
the Eguchi-Hanson and following the discussion in section 2, the
smooth pp-wave limit  of $AdS_5\times \RP^5$ is
\be
ds^2=-4 dx^+dx^--\mu^2\left({\vec{x}}{~}^2+r^2+{a^2\over 2}
\ln\left(a-r\over a+r\right)\right)){dx^+}^2+d{\vec{x}}{~}^2 +
ds_{EH}^2
\ee
where $ds_{EH}^2$ is the Eguchi-Hanson metric.

Next, we will consider the near horizon limit of $N$ D3-branes on an
orientifold O7-plane. We will also put D7-branes sitting together with
the D3's so that the
dilaton is constant and the low-energy theory is conformal 
\cite{Sen:1996vd},\cite{Banks:1996nj}, \cite{Dasgupta:1996ij} .
At generic point in the D7-brane moduli space,
we have non-conformal field theories living on the D3-branes.
There exist 7 different types 
of singularities which give rise to constant dilaton.
These are the Argyres-Douglas points $H_0,~ H_1$ \cite{arg},\cite{argwit} and
$H_2$ with
$A_0,~A_1,$ and $A_2$ gauge theories on the D7 and $D_4$, $E_6$,
$E_7$ and $E_8$
theories resulting from corresponding singularities in F-theory.
Among these singularities, only the $D_4$
can occur for any value of the dilaton. 
All the others appear at fixed, order one, string coupling.
 The
resulting field theory of coincident $N$ D3 and D7-branes at
a $D_4$ singularity
is ${\cal{N}}=2$ $USp(2N)$ SYM theory with
a hypermultiplet in the anti-symmetric representation and four
hypermultiplets in the fundamental. The $\Z_2$
action has now fixed  points on the $S^5$. In fact, the fixed point
set (where both the O7 and the D7's are) is $AdS_5\times S^3$.
The supergravity description of D3-branes near D7-brane
singularities has been described in \cite{Fayyazuddin:1998fb}
(for $D_4,E_6,E_7,E_8$)
and the metric can be written as
\be
&&R^{-2}ds^2=d\rho^2+\sinh^2\!\rho \, d\Omega_3^2-
\cosh^2\!\rho\,
dt^2+d\theta^2+\sin^2\!\theta\,
d\psi^2+\nonumber \\
&&\phantom{xxxxxxxx}\cos^2\!\theta\left(d\omega^2+\sin^2\!\omega\,
d\chi^2+\cos^2\! \omega\, d\phi^2\right)
\, ,\label{orr}
\ee
where $0\leq \theta<\pi/2$ and $\phi$ is periodic with period
$2\pi(1-\alpha/2)$. The values of $\alpha$ are $\alpha={1\over
3},~
{1\over 2},~ {2\over 3},~ 1,~ {4\over 3},~ {3\over 2},~ {5\over
3}$, for $A_0,~ A_1,~A_2,~D_4,~E_6, ~E_7$ and $E_8$ 
\cite{Aharony:1998xz}. The B-fields
have generally $SL(2,\Z)$ monodromies around the $\phi$-circle.

We may now consider the standard Penrose limit as in
eq.(\ref{pen}). For this we may first replace $\theta$ with
$\theta-\pi/2$ and then take the limit $R\to \infty$ after the
transformation eq.(\ref{pen5}). However, this limit leads to the
 background  in eq.(\ref{pp4}).  Moreover, the monodromies of the
B-fields are washed out. However, there exist another limit which preserves
the singularities. Defining the coordinates
\be
x^+={t+\psi\over 2}\, , ~~~x^-=R^2\left({t-\psi\over 2}\right)\, , ~~~
x=R\, \rho\, , ~~~ y=R\, \theta\, ,~~~w= R\, \omega \label{pen55}
\ee
and taking  $R\to \infty $, we end up with
the pp-wave limit of (\ref{orr})
\be
&&ds^2=-4dx^+dx^--\mu^2(x^2\!+\!y^2\!+\! w^2){dx^+}^2\!+\!dx^2+x^2d\Omega_3^2+\nonumber
\\&&\phantom{xxxxxxx}
dy^2+y^2 d\chi^2+dw^2+w^2d\phi^2\,. \label{pprs}
\ee
This can be written as
\be
&&ds^2=-4dx^+dx^--\mu^2\left(\vec{r}{~}^2\!+
\!{|z|^{2-\alpha}\over (1-{\alpha\over 2})^2}\right){dx^+}^2\!+
\!d\vec{r}{~}^2+{|dz|^2\over |z|^\alpha}\,. \label{pprss}
\ee
where
$\vec{r}$ is in $\R^6$. Clearly then, the wave-front geometry is
$\Co^3\times \Co/\Z_{n}$ where $n=2,3,4,6$  for D3-branes near
$D_4,E_6,E_7$ and $E_8$ singularities in F-theory.
 
\section{Conclusions}

We  presented above various orbifolds of $AdS_3\times S^3$,
$AdS_5\times S^5$ and $AdS_{4,7}\times S^{7,4}$ as well as
orientifolds of $AdS_5\times S^5$. The limiting pp-waves are in
general singular. Some of them can be desingularized by replacing
their wave fronts (where the singularities are located) by Ricci flat
spaces of appropriate holonomy in order to preserve
supersymmetry. There exist two ways to view these orbifolds, either as
the limiting spaces of $AdS_p\times S^q/\Gamma$ 
orbifolds,  or as orbifolds of the pp-wave limit of  $AdS_p\times
S^q$. In the latter case, we recall from appendix A that, 
in general,  a D-dimensional pp-wave background can be embedded  in
$M^{2,D}$ flat space-time as
\be
&&
X_D-X_{D+1}=\left(X_0+X_{D-1}\right)\nonumber \\
&&X_D+X_{D+1}= S(X^i)\, , ~~~~~i=1,...,D-2\, .
\ee
We may then consider a discrete action of $\Gamma\subset SO(D-2)$  on
 the wave-front
coordinates $X^i$. Depending on the geometry of the wave fronts,
these orbifolds may or may not preserve supersymmetry. The
orbifolds we have considered here are all  supersymmetric.

\appendix

\section{Embedding the pp-waves in higher dimensions}
\vskip .2in

$AdS_5\times S^5$ which can be embedded in the flat 12D
space-time $M^{2,10}$ of $(-,+,+,...+,-)$ signature and metric
\be
ds_{12}^2=-dX_0^2+dX_1^2+....+dX_{10}^2-dX_{11}^2 \label{mmM}
\ee
Similarly, it is straightforward  to show that the pp-wave background
\be
&&ds^2=-4dx^+dx^--\mu^2{\vec{r}}{~}^2{dx^+}^2+d{\vec{r}}{~}^2
\label{ppp}
 \ee
can be embedded in $M^{2,10}$
as the hypersurface
\be
&&X_{10}-X_{11}=\left(X_0+X_{9}\right)^2\, ,\\
&&X_{10}+X_{11}={\mu^2\over 8}\left(X_1^2+X_2^2+...+X_8^2\right)
\ee
Parametrizing the above hypersurface by
\be
&&X_1=x_1\,, ~~~X_2=x_2\, , ~~~...~~~X_8=x_8\, , \\
&&X_0=\mu^2\left(x_1^2+x_2^2+...+x_8^2\right){x^+\over 4}+x^-+x^+\,
, \\&&
X_{9}=-\mu^2\left(x_1^2+x_2^2+...+x_8^2\right){x^+\over 4}-x^-+x^+\,
, \\&&
X_{10}={\mu^2\over 16}\left(x_1^2+x_2^2+...+x_8^2\right)+
2x^+{}^2\, , \\&&
X_{11}={\mu^2\over 16}\left(x_1^2+x_2^2+...+x_8^2\right)-
2x^+{}^2\, ,
\ee
leads to the pp wave metric (\ref{ppp}).

More general, the D-dimensional pp-wave metric
\be
&&ds^2=-4dx^+dx^--8 S(x^i){dx^+}^2+d{\vec{r}}{~}^2\, , ~~~i=1,...,D-2
 \ee
can be embedded in the (D+2)-dimensional
 flat space time $M^{2,D}$ with metric 
\be
ds_{12}^2=-dX_0^2+dX_1^2+....+dX_{D}^2-dX_{D+1}^2\, ,  \label{mmM}
\ee
as
\be
&&X_{D}-X_{D+1}=\left(X_0+X_{D-1}\right)^2\, ,\\
&&X_{D}+X_{D+1}=S(X^i)\, , 
\ee
which can be parametrized as
\be
&&X_1=x_1\,, ~~~X_2=x_2\, , ~~~...~~~X_{D-2}=x_{D-2}\, , \\
&&X_0=2 S x^++x^-+x^+\,
, ~~~
X_{D-1}=-2 S x^+-x^-+x^+\,
, \\&&
X_{D}={S\over 2}+
2x^+{}^2\, , ~~~
X_{D+1}={S\over 2}-
2x^+{}^2\, .
\ee
Emmbeding of the pp-wave as intersection of quadratic
surfaces in  $M^{2,D}$ has  been discussed in \cite{Blau:2002rg}. 
\vskip .2in

\vskip .2in

\section{A desingularized  $\Co^3/\Z_3$ space}

We will describe the metric of a non-compact 6D space with
holonomy $SU(3)$ which is asymptotically $\Co^3/\Z_3$.
For this, we start by considering the five-dimensional sphere  $S^5$, which
is the surface
$$
Z_1\bar Z_1+Z_2\bar Z_2+Z_3\bar Z_3=1 \ ,
$$
embedded in  $\Co^3$. By expressing the $Z$'s as 
 \cite{Kehagias:2000dg}
\be
Z_1\!=\!{ e^{i\t}\over (1+{r^2})^{1/2}}\ ,~~ Z_2\!=\!{r 
e^{i(\t+{\chi+\varphi\over 2})} \over
(1+{r^2})^{1/2}}\!\sin{\theta\over 2}
\, , ~~
Z_3\!=\!{r e^{i(\t+{\chi-\varphi\over
2})}\over (1+{r^2})^{1/2}}\cos{\theta\over 2}\, ,  \label{Z}
\ee
the metric of the $S^5$ turns out to be 
\be
ds^2(S^5)=d\Sigma_4^2 +(d\t+A)^2\ , \ee where
\be
d\Sigma_4^2={dr^2\over (1+{r^2})^2}+ {1\over 4}{r^2\over
(1+{r^2})}
\left(d\theta^2+\sin^2\theta d\varphi^2\right)+{1\over 4}{r^2\over
(1+{r^2})^2}\left(d\chi-\cos\theta d\varphi\right)^2 \, ,
\label{s4}
\ee
\be
A={r^2\over 2 ( 1+ r^2) }\big(d\chi -\cos\theta
d\varphi\big) \ .
\ee
The metric  $d\Sigma_4$ is just the metric of $\CP^2$ and
 eq.(\ref{s4}) makes manifest the Hopf-fibration of $S^5$, that is a $U(1)$
bundle over $\CP^2$. We may search for $SU(3)$-holonomy manifolds 
by making the ansatz
\be
d\sigma^2=f(r)dr^2+r^2d\Sigma_4^2+r^2h(r)(d\t+A)^2\ ,
\ee
where $r$ is a radial coordinate and $f(r),h(r)$ are functions of $r$ which
will be determined by demanding the manifold to be Ricci flat. Indeed,
the Ricci-flatness condition for this geometry, turns out to be, 
for $f=1/h$, the single equation
\be
r^2 h h''+3r  h h'+r^2 {h'}^2 -4h^2=0\, .
\ee
The asymptotically locally flat solution of this equation is
\be
h(r)=1-{L^6\over r^6}
\ee
and the metric is
\be
d\sigma^2={dr^2\over1-{L^6\over r^6}} +r^2d\Sigma_4^2+r^2
(1-{L^6\over r^6})(d\t+A)^2\ . \label{SS1}
\ee
It seems that a singularity exist at $r=L$, which
 as one suspects can be cured. 
At $r=L$ we get
\be
ds^2\sim {1\over 9}\left(du^2+9\ u^2(d\t+A)^2\right)+L^2d\Sigma_4^2\, 
\ee
and there exist a conical singularity, which  
is removed if the periodicity of $\tau$ is not
$2\pi$ as it seems from (\ref{Z}),
but rather $2\pi/3$.
As a  result, although at $r\to \infty$ the space is flat, it is not
globally flat in view of the $\tau$ periodicity. In fact it is an ALE space,
asymptotically $\Co^3/\Z_3$. That it is of $SU(3)$ holonomy can also
be proven without much effort. 
\vskip .1in

{\bf Note Added} During the completion of this work, we became aware
of the works  \cite{Alishahiha:2002ev}, \cite{Kim:2002fp} and 
\cite{Takayanagi:2002hv} where superstring on 
$AdS_5\times S^5/\Z_N$ orbifolds are studied.
\vskip .1in

\acknowledgments

This work is partially supported by the European RTN networks
HPRN-CT-2000-00122 and HPRN-CT-2000-00131.


\vskip .2in



\begin{thebibliography}{999}

\bibitem{Amati:1988ww}
D.~Amati and C.~Klimcik,
``Strings In A Shock Wave Background And Generation Of Curved 
Geometry From Flat Space String Theory,''
Phys.\ Lett.\ B {\bf 210}, 92 (1988).

\bibitem{Horowitz:bv}
G.~T.~Horowitz and A.~R.~Steif,
``Space-Time Singularities In String Theory,''
Phys.\ Rev.\ Lett.\  {\bf 64}, 260 (1990).

\bibitem{Blau:2001ne}
J.~Figueroa-O'Farrill and G.~Papadopoulos, ``Homogeneous fluxes, 
branes and a maximally supersymmetric solution
of  M-theory,'' JHEP {\bf 0108}, 036 (2001), hep-th/0105308;\\ 
M.~Blau, J.~Figueroa-O'Farrill, C.~Hull and G.~Papadopoulos, ``A new 
maximally supersymmetric background of IIB superstring theory,''
JHEP {\bf 0201}, 047 (2002), hep-th/0110242.

\bibitem{Blau:2002dy}
M.~Blau, J.~Figueroa-O'Farrill, C.~Hull and G.~Papadopoulos,
``Penrose limits and maximal supersymmetry,'', hep-th/0201081.

\bibitem{Blau:2002rg}
M.~Blau, J.~Figueroa-O'Farrill and G.~Papadopoulos,
``Penrose limits, supergravity and brane dynamics,''
hep-th/0202111.


\bibitem{Kowalski-Glikman:wv} J. Kowalski-Glikman,
``Vacuum States In Supersymmetric Kaluza-Klein Theory,'' Phys.
Lett. B {\bf 134}, 194 (1984).


\bibitem{Penrose}
R. Penrose, ``Any spacetime has a plane wave as a limit'',
Differential geometry and relatrivity, Reidel, Dordrecht, 1976. 

\bibitem{Gueven:2000ru}
R.~Gueven,
``Plane wave limits and T-duality,''
Phys.\ Lett.\ B {\bf 482}, 255 (2000), hep-th/0005061.

\bibitem{Berenstein:2002jq}
D.~Berenstein, J.~Maldacena and H.~Nastase,
``Strings in flat space and pp waves from N = 4 super Yang Mills,''
hep-th/0202021.


\bibitem{Nahm:1977tg}
W.~Nahm,
``Supersymmetries And Their Representations,''
Nucl.\ Phys.\ B {\bf 135}, 149 (1978).

\bibitem{Hatsuda:2002xp}
M.~Hatsuda, K.~Kamimura and M.~Sakaguchi,
``From super-$AdS(5) \times S^5$ algebra to super-pp-wave algebra,''
hep-th/0202190.

\bibitem{Sen:1996vd}
A.~Sen,
``F-theory and Orientifolds,''
Nucl.\ Phys.\ B {\bf 475}, 562 (1996), hep-th/9605150.

\bibitem{Banks:1996nj}
T.~Banks, M.~R.~Douglas and N.~Seiberg,
``Probing F-theory with branes,''
Phys.\ Lett.\ B {\bf 387}, 278 (1996),  hep-th/9605199 .

\bibitem{Dasgupta:1996ij}
K.~Dasgupta and S.~Mukhi,
``F-theory at constant coupling,''
Phys.\ Lett.\ B {\bf 385}, 125 (1996), hep-th/9606044.

\bibitem{arg}P.C. Argyres and M.R. Douglas, ``New Phenomena in $SU(3)$
Supersymmetric Gauge Theory,'' Nucl. Phys. B {\bf 448}, 93 (1995), 
hep-th/9505062.  

\bibitem{argwit} P.C. Argyres, M.R. Plesser, N. Seiberg and E. Witten,
``New $N=2$ Superconformal Field Theories in Four Dimensions,''
Nucl. Phys. B {\bf 461}, 71 (1996), hep-th/9511154.


\bibitem{Aharony:1999ti}
O.~Aharony, S.~S.~Gubser, J.~Maldacena, H.~Ooguri and Y.~Oz,
``Large N field theories, string theory and gravity,''
Phys.\ Rept.\  {\bf 323}, 183 (2000), hep-th/9905111.

\bibitem{Horowitz:1996ay}
G.~T.~Horowitz, J.~M.~Maldacena and A.~Strominger,
``Nonextremal Black Hole Microstates and U-duality,''
Phys.\ Lett.\ B {\bf 383}, 151 (1996), hep-th/9603109.


\bibitem{Maldacena:1997re}
J.~Maldacena,
``The large $N$ limit of superconformal field theories and supergravity,''
Adv.\ Theor.\ Math.\ Phys.\  {\bf 2}, 231 (1998)
[Int.\ J.\ Theor.\ Phys.\  {\bf 38}, 1113 (1998)], hep-th/9711200.

\bibitem{Metsaev:2001bj}
R.~R.~Metsaev,
``Type IIB Green-Schwarz superstring in plane wave Ramond-Ramond  
background,''
Nucl.\ Phys.\ B {\bf 625}, 70 (2002), hep-th/0112044.


\bibitem{Metsaev:2002re}
R.~R.~Metsaev and A.~A.~Tseytlin,
``Exactly solvable model of 
superstring in plane wave Ramond-Ramond  background,''
hep-th/0202109.

\bibitem{Russo:2002rq}
J.~G.~Russo and A.~A.~Tseytlin,
``On solvable models of type IIB superstring in 
NS-NS and R-R plane wave  backgrounds,'' hep-th/0202179.

\bibitem{Kehagias:1994iy}
A.~A.~Kehagias and P.~A.~Meessen,
``Exact string background from a WZW model based on the Heisenberg group,''
Phys.\ Lett.\ B {\bf 331}, 77 (1994), hep-th/9403041;
A.~A.~Kehagias,
``All WZW models in D$\leq$ 5,'' hep-th/9406136.

\bibitem{Nappi:1993ie}
C.~R.~Nappi and E.~Witten,
``A WZW model based on a nonsemisimple group,''
Phys.\ Rev.\ Lett.\  {\bf 71}, 3751 (1993), 
hep-th/9310112.

\bibitem{Sfetsos:1994vz}
K.~Sfetsos,
``Gauged WZW models and nonAbelian duality,''
Phys.\ Rev.\ D {\bf 50}, 2784 (1994), hep-th/9402031.

\bibitem{Kehagias:1998gn}
A.~Kehagias,
``New type IIB vacua and their F-theory interpretation,''
Phys.\ Lett.\ B {\bf 435}, 337 (1998), hep-th/9805131.


\bibitem{Klebanov:1998hh}
I.~R.~Klebanov and E.~Witten,
``Superconformal field theory on threebranes at a Calabi-Yau  singularity,''
Nucl.\ Phys.\ B {\bf 536}, 199 (1998), hep-th/9807080.

\bibitem{Morrison:1998cs}
D.~R.~Morrison and M.~R.~Plesser,
``Non-spherical horizons. I,''
Adv.\ Theor.\ Math.\ Phys.\  {\bf 3}, 1 (1999), hep-th/9810201.


\bibitem{Douglas:1996sw}
M.~R.~Douglas and G.~W.~Moore,
``D-branes, Quivers, and ALE Instantons,''
hep-th/9603167.

\bibitem{Douglas:1997de}
M.~R.~Douglas, B.~R.~Greene and D.~R.~Morrison,
``Orbifold resolution by D-branes,''
Nucl.\ Phys.\ B {\bf 506}, 84 (1997), hep-th/9704151.

\bibitem{Kachru:1998ys}
S.~Kachru and E.~Silverstein,
``4d conformal theories and strings on orbifolds,''
Phys.\ Rev.\ Lett.\  {\bf 80}, 4855 (1998), hep-th/9802183.

\bibitem{Itzhaki:2002kh}
N.~Itzhaki, I.~R.~Klebanov and S.~Mukhi,
``PP wave limit and enhanced supersymmetry in gauge theories,'' hep-th/0202153.

\bibitem{Gomis:2002km}
J.~Gomis and H.~Ooguri,
``Penrose limit of N = 1 gauge theories,''
hep-th/0202157.

\bibitem{Zayas:2002rx}
L.~A.~Zayas and J.~Sonnenschein,
``On Penrose limits and gauge theories,''
hep-th/0202186.


\bibitem{Horowitz:1998xk}
G.~T.~Horowitz and D.~Marolf,
``A new approach to string cosmology,''
JHEP {\bf 9807}, 014 (1998), hep-th/9805207.

\bibitem{Gao:1999er}
Y.~H.~Gao,
``AdS/CFT correspondence and quotient space geometry,''
JHEP {\bf 9904}, 005 (1999), hep-th/9903080.

\bibitem{Lawrence:1998ja}
A.~E.~Lawrence, N.~Nekrasov and C.~Vafa,
``On conformal field theories in four dimensions,''
Nucl.\ Phys.\ B {\bf 533}, 199 (1998), hep-th/9803015.

\bibitem{Morrison} D.R. Morrison and G. Stevens, ``Terminal quotient
singularities in dimensions three and four'', Math. Comp. {\bf 51}
(1984) 15.

\bibitem{Ferrara:1998vf}
S.~Ferrara, A.~Kehagias, H.~Partouche and A.~Zaffaroni,
``Membranes and fivebranes with lower
 supersymmetry and their AdS  supergravity duals,''
Phys.\ Lett.\ B {\bf 431}, 42 (1998), hep-th/9803109.

\bibitem{Giveon:1998sr}
A.~Giveon and D.~Kutasov,
``Brane dynamics and gauge theory,''
Rev.\ Mod.\ Phys.\  {\bf 71}, 983 (1999), hep-th/9802067.

\bibitem{Witten:1998xy}
E.~Witten,
``Baryons and branes in anti de Sitter space,''
JHEP {\bf 9807}, 006 (1998), hep-th/9805112.

\bibitem{Aharony:1998xz}
O.~Aharony, A.~Fayyazuddin and J.~Maldacena,
``The large N limit of N = 2,1 field theories from three-branes in  F-theory,''
JHEP {\bf 9807}, 013 (1998), hep-th/9806159.



\bibitem{Fayyazuddin:1998fb}
A.~Fayyazuddin and M.~Spalinski,
``Large N superconformal gauge theories and supergravity orientifolds,''
Nucl.\ Phys.\ B {\bf 535}, 219 (1998), hep-th/9805096.

\bibitem{Kehagias:2000dg}
A.~Kehagias and J.~G.~Russo,
``Hyperbolic spaces in string and M-theory,''
JHEP {\bf 0007}, 027 (2000), hep-th/0003281.

\bibitem{Alishahiha:2002ev}
M.~Alishahiha and M.~M.~Sheikh-Jabbari,
``The PP-wave limits of orbifolded $AdS(5) \times S^5$,'' hep-th/0203018.


\bibitem{Kim:2002fp}
N.~Kim, A.~Pankiewicz, S.~J.~Rey and S.~Theisen,
``Superstring on PP-Wave Orbifold from Large-N Quiver Gauge Theory,''
hep-th/0203080.

\bibitem{Takayanagi:2002hv}
T.~Takayanagi and S.~Terashima,
``Strings on Orbifolded PP-waves,''
hep-th/0203093.

\end{thebibliography}
\end{document}